\documentclass[letterpaper]{sfchem}
\usepackage{graphicx}

\begin{document}

\title{Does IRAS 16293--2422 contain a hot core? New interferometric results}

\author{Fredrik L.\ Sch\"oier\inst{1} \and Jes K. J{\o}rgensen\inst{1} \and Ewine F.\ van Dishoeck\inst{1} \and Geoffrey A.\ Blake\inst{2}} 
  \institute{Leiden Observatory,  
  P.O.\ Box 9513, 2300 RA Leiden, The Netherlands 
  \and Division of Geological and Planetary Sciences, California Institute of Technology, 
  Pasadena, CA  91125, USA
} 
\authorrunning{Sch\"oier et al.}
\titlerunning{Does IRAS 16293--2422 contain a hot core?}

\maketitle 

\begin{abstract}
High angular resolution H$_2$CO line observations have been carried out for the low mass proto-binary star IRAS 16293--2422 using the OVRO millimeter array. Simultaneous continuum observations reveal that the source most probably contains two accretion
disks, less than $\sim 250$\,AU in size, through which matter is fed onto the stars.
The observations suggest that the binary has cleared  most  of the material in the inner part of the envelope, out to the binary separation.
The H$_2$CO interferometer observations indicate the presence of large scale rotation roughly perpendicular to the large scale CO outflow associated with this 
source. The H$_2$CO emission is dominated by the compact dense and hot ($T>100$\,K) gas close to the positions of the continuum peaks. 


\end{abstract}

\section{Introduction}
Recent studies (J{\o}rgensen et~al.\ 2002; Sch\"oier et~al.\ 2002; Shirley et~al.\ 2002) have suggested that the inner envelopes of low-mass protostars are dense ($>10^6$\,cm$^{-3}$) and warm ($>80$\,K), as would be expected from scaling of high-mass protostars (Ceccarelli et~al.\ 1996; Ivezi{\' c} \& Elitzur 1997). 
Detailed modelling of multi-transition single-dish observations towards the deeply embedded low mass protostellar object  IRAS 16293--2422
(Ceccarelli et~al.\ 2000; Sch\"oier et~al.\ 2002) shows that the abundances of some molecules, e.g. H$_2$CO and CH$_3$OH, are drastically increased in the warm and dense inner 
region of the circumstellar envelope. 
 The location where this increase occurs is consistent with the radius at which ices are expected to thermally evaporate 
off the grains ($\sim 90$\,K). 
However, the small spatial scale of the region of hot gas ($<100$\,AU) and the infalling nature of the envelope lead to very different chemical time scales between low mass and high mass hot cores, which may prevent production of second-generation complex molecules in low mass protostars.
 Also, shocks are thought to be relatively more important in low mass protostars. 
High angular resolution observations are needed to pinpoint the origin of the abundance enhancements and possibly distinguish these two scenarios.
We present here observations of H$_2$CO at 1\,mm using the Owens Valley Radio Observatory (OVRO) millimeter array at $\sim 3\arcsec$ resolution.
The line ratio of the two H$_2$CO lines observed is a measure of the gas temperature.
  
\section{Observations and data reduction}
IRAS 16293--2422 was observed using the OVRO millimeter array
between September 2000 and March 2002. The continuum emission at 1.37\,mm was obtained simultaneously with
the H$_2$CO $3_{22}\rightarrow 2_{21}$ and $3_{03}\rightarrow 2_{02}$ line emission at 218.222 and 218.475\,GHz, respectively.
\object{IRAS 16293--2422} was observed in the L and E configurations. The natural weighted
continuum observations has a 1$\sigma$ noise of about
20\,mJy\,beam$^{-1}$ with a beam size of 3$\farcs$9$\times$1$\farcs$9.

\section{Results and discussion}

\begin{figure} 
\centering{\includegraphics[width=88mm]{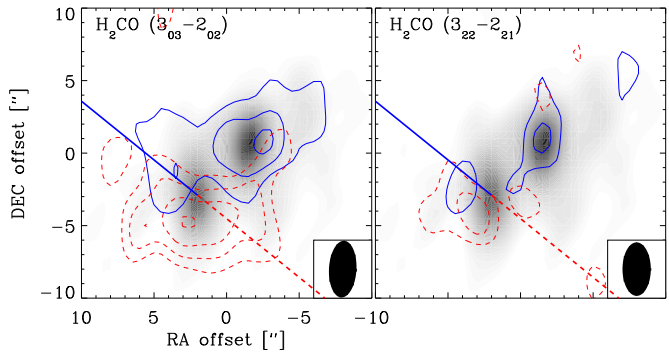}
\caption{OVRO interferometer maps of H$_2$CO emission (contours)  overlayed on the 1.37\,mm 
continuum emission (greyscale) from IRAS 16293--2422. The H$_2$CO emission has been separated into a blue (4-7\,km\,s$^{-1}$; solid lines)  and a red (1-4\,km\,s$^{-1}$; dashed lines) part. Contours start at the 2$\sigma$-level (1.0\,Jy\,beam$^{-1}$\,km\,s$^{-1}$) with contours spaced by 2$\sigma$. 
Also indicated is the direction of the large scale CO outflow. The southeast continuum source is MM1 whereas MM2 is to the northwest.}  
\label{h2co}}
\end{figure}

\begin{figure} 
\centering{\includegraphics[width=8cm]{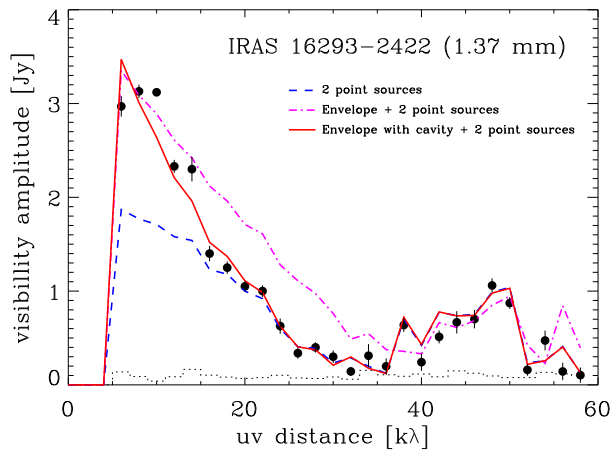}
\caption{Visibility amplitudes of the OVRO 1.37\,mm continuum emission towards IRAS 16293--2422
as function of the projected baseline length. The central position is taken to be midway between MM1 and MM2. The observations are plotted 
as filled symbols with 1$\sigma$ error bars. 
Also shown are the predictions from the continuum model (Sch\"oier et al.\ 2002), with the same $uv$ sampling as the observations (dash-dotted magenta line)
with unresolved compact emission (dashed blue line) added to the model. 
A model envelope with a cavity (solid red line) is shown to best reproduce the observations.}  
\label{cont_vis}}
\end{figure}
\begin{figure} 
\centering{\includegraphics[width=60mm]{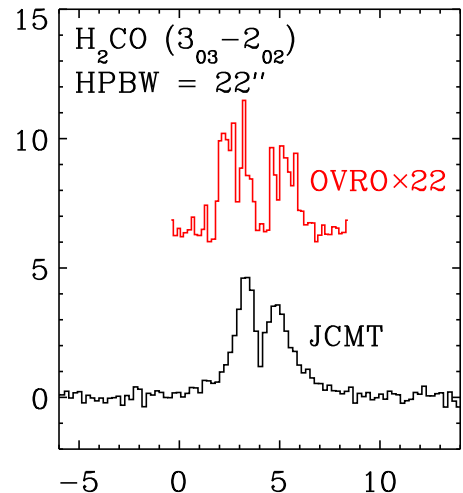}}
\caption{Comparison between the H$_2$CO ($3_{03}\rightarrow 2_{02}$) line emission towards \object{IRAS 16293--2422} at the source position from JCMT single-dish observations (lower spectrum) and OVRO interferometric observations (upper spectrum) restored with the JCMT beam (22\arcsec). The OVRO spectrum has been scaled by a factor of 22 in order to account for all the flux seen in the JCMT spectrum.}
\label{compare}
\end{figure}

\subsection{Continuum emission}
In Fig.~\ref{h2co}  the 221.7\,GHz (1.37\,mm) continuum emission observed toward IRAS 16293--2422 is presented.
As expected for this proto-binary star, two continuum sources are detected, separated by approximately $5\arcsec$ 
with faint extended emission also present. The total observed continuum flux density at 1.37\,mm  is about 3.5\,Jy. This is roughly 50\% of the flux 
obtained by Walker et~al.\ (1990) from mapping with a single dish telescope, indicating that the interferometer resolves out some of the emission. 
The positions of the continuum sources are consistent with the two 3\,mm sources MM1 (southeast) and MM2 (northwest) found by Mundy et~al.\ (1992).  
At the distance of IRAS 16293--2422 (160\,pc) the projected separation of the sources is about $\sim 800$\,AU. 
The sources are unresolved and the fitted sizes in the $uv$-plane of MM1 and MM2 provide upper limits to the sizes of these disks of $\sim 250$\,AU in diameter.
The total fluxes estimated for the compact components MM1 and MM2 are 0.7\,Jy and 1.1\,Jy, respectively. The remaining 1.7\,Jy of emission is attributed to the contribution from an extended circumbinary envelope. Assuming the emission to be optically thin and adopting a dust temperature of 40\,K sets lower limits to the masses of 0.25 and 0.4\,M$_{\odot}$, respectively.

Sch\"oier et al.\ (2002) determined the circumstellar structure of  IRAS 16293--2422 from detailed modelling of the observed continuum emission. In addition to 
the SED, resolved images at 450 and 850\,$\mu$m obtained with the SCUBA bolometer array at the JCMT were used to constrain the large scale envelope structure. In order to investigate if the brightness distribution at 1.37\,mm from the envelope model is consistent with the flux picked up by the interferometer, the same $uv$ sampling was applied to the model output. Furthermore, the interferometer data constrain the envelope structure at smaller scales ($\sim 1\arcsec$) than the JCMT single-dish data ($\sim 10\arcsec$). In Fig.~\ref{cont_vis} the visibility amplitudes of the
observed emission are compared to the model predictions. Fitting just two compact sources underestimates the flux at small baselines, indicating additional large scale emission. Addition of the best fit envelope model of Sch\"oier et al.\ (2002) produces the correct amount of flux at the smaller baselines but provides too much emission at intermediate baselines ($\sim 5\arcsec$), i.e., at scales of the binary separation. It is found that an envelope model that is void of material on scales smaller than the binary separation best reproduces the observed visibilities. The emission from the two unresolved components, possibly accretion disks, is estimated to contribute $\sim 25$\% to the total flux at 1.37\,mm. 

\subsection{Molecular line emission}
The observed H$_2$CO ($3_{22}\rightarrow 2_{21}$ and $3_{03}\rightarrow 2_{02}$) line emission obtained at OVRO is shown in Fig.~\ref{h2co}.
The emission is clearly resolved and shows a $\sim 6\arcsec$ separation between the red ($4-7$\,km\,s$^{-1}$) and blue ($1-4$\,km\,s$^{-1}$) 
emission peaks. The direction of the red-blue asymmetry is roughly perpendicular to the large scale CO outflow associated with MM1
(Walker et~al.\ 1988) and is indicative of an overall rotation of the material encompassing both MM1 and MM2.  
The emission from both lines peaks close to the positions of MM1 and MM2. 
The ($3_{22}\rightarrow 2_{21}$)/($3_{03}\rightarrow 2_{02}$) line ratio is sensitive to the temperature of the emitting gas (e.g., van Dishoeck et~al.\ 1995), and is $> 0.4$ close to the positions of MM1 and MM2 indicating hot ($>100$\,K) gas.  Single-dish observations (van Dishoeck et~al.\ 1995; Sch\"oier et~al.\ 2003, in prep.) show that the H$_2$CO line emission is extended to scales of $\sim 30\arcsec$.
Also, the large scale ($3_{22}\rightarrow 2_{21}$)/($3_{03}\rightarrow 2_{02}$) line ratio is $\sim 0.25$, much lower than for the interferometer data, suggesting that a cold ($<50$\,K) component dominates the single-dish flux.

Care has to be taken when interpreting the interferometer maps due to the low sensitivity to weak large scale emission. A direct comparison between the 
single-dish spectrum and that obtained from the interferometer observations restored with the single-dish beam is shown in Fig.~\ref{compare}. The interferometer  picks up only $\sim 5$\% of the single dish flux, suggesting that the extended cold material is resolved out by the interferometer and that a hotter compact component is predominantly picked up.
This is to be expected from the `jump models' introduced by Sch\"oier et al.\ (2002) where a steep gradient in abundance  needed to be introduced, for a
number of molecular species including H$_2$CO,  in order to model emission lines probing a wide range of excitation conditions.
Simulations where the $uv$ sampling from the observations has been applied to the best fit H$_2$CO models from Sch\"oier et al.\ (2002) with and without a
jump in the abundance show that the `hot core' component dominates the flux as seen by the interferometer ($>50$\%).

\section{Summary and outlook}
Interferometric continuum observations suggest that the inner parts of the circumbinary envelope around  IRAS 16293--2422 is relatively void of material
on scales smaller than the binary separation ($\sim 5\arcsec$). The simultaneous H$_2$CO line observations indicate presence of hot and dense gas close to the peak positions of the continuum emission consistent with thermal evaporation of ices in `hot core' regions close to the protostars. Based on the morphology and line widths  little of the observed interferometric emission is
thought to be associated with the known outflow(s) associated with this source.

Additional interferometric data have been obtained for  IRAS 16293--2422 in a wide range of lines including C$^{18}$O,  CH$_3$OH, HCN, N$_2$H$^{+}$ and SiO, that will provide additional constraints on the physical and chemical structure in the circumstellar envelope.
 Whatever their precise origin, the molecules located in the inner
envelope of \object{IRAS 16293--2422} can be incorporated into the
growing circumstellar disk(s) and become part of the material from
which planetary bodies are formed.

\begin{acknowledgements}
This research was supported by the Netherlands Organization for
Scientific Research (NWO) grant 614.041.004, the Netherlands Research School
for Astronomy (NOVA) and a NWO Spinoza grant.
OVRO observations were supported by the National Science Foundation, grant AST-9981546.
\end{acknowledgements}

\end{document}